# Review of Machine Learning Applications in Wireless Communications

Apoorva Subhash Bajaj
*Dept. of Electrical and Electronic Engineering*
*Birla Institute of Technology and Science, Pilani - K.K Birla Goa Campus*
Goa, India
f20180269@goa.bits-pilani.ac.in

*Abstract*—This paper looks at various aspects of Machine Learning (ML) applications in wireless communication technologies, focusing mainly on fifth-generation (5G) and millimeter wave (mmWave) technologies. This paper includes the summaries of 3 papers on machine learning applications in wireless communication technology. The paper deals with the need for integration of machine learning in wireless communication, types of machine learning techniques used in wireless communication, advantages and potential of ML in wireless communication, and implementation parameters of ML in wireless communication, as well as a study on RSS-Based Classification of usage in indoor millimeter-wave wireless networks.

## I. INTRODUCTION

The next generation of wireless communication networks are becoming more and more complex due to a wide number of service requirements, diverse characteristics of applications in the industry, and devices themselves. Traditional networking approaches, i.e., reactive, centrally-managed, one-size-fits-all approaches, and conventional data analysis tools, have limited capacity [1]. These approaches do not support future complex wireless networks with regards to the operation and optimization, and cost-efficient demands of the networks and network providers. The exponentially increasing demand in the number of people using wireless communication technologies requires innovation and research into new technologies and optimization techniques. Machine learning (ML) is emerging as one of the most promising technologies that will help engineers conceptualize and implement technologies like autonomous vehicles, industrial and home automation, virtual and augmented reality, e-health, and many more such applications. The authors of paper [1] lay out the factors that are driving the use of ML into wireless communication technologies, including the need to move away from traditional optimization techniques and exploit Big Data to cost-efficiently provide the best quality of service (QoS) to users and managing complexity of future networks. The authors of paper [2] layout the different ML Models that are used to implement new wireless communication technologies while reviewing several papers to provide a conclusive study of the use of ML in optimization. They also discuss several benefits and applications of ML in wireless communication technologies with a focus on the fifth-generation (5G) wireless communication. As mmWave is a technology which aims to free up congested bandwidth at the ultra-high frequency range of 300 MHz to 3 GHz, and is also an integral part of 5G in providing data speeds of up to 100 Gb/s, we also discuss a use case of Machine learning-based optimization of Indoor Millimeter Wave (mmWave) networks.

## II. NEED OF MACHINE LEARNING IN WIRELESS COMMUNICATION

The paper [1] discusses the need for ML in wireless communication technologies along with the factors driving it. The traditional and prevailing approach involves the independent optimization of single key performing indicators (KPI). This means that the engineers are using a limited set of data belonging only to the network to carry out data analysis and optimize the network. The paper [1] suggests the value brought to optimization by Data Analytics is only beneficial when the range of data sources is expanded, and a user-centric, quality of experience (QoE) approach is adopted to optimize end-to-end network performance and efficiency. There are multiple factors that are driving network operators and designers to implement ML. The paper divides the factors into three categories namely *Cost and Service*, *Usage*, and *Technological* drivers.

The demand of users is increasing exponentially, but users are not increasing their wireless payout, which generates an urgent need for cost-efficient optimization techniques. The paper suggests that network operators need to manage traffic based on QoS, improve efficiency to retain users while maintaining a stable profit margin, improve the performance of the network, and QoE.

The next generation of wireless networks needs a robust analytics framework to provide virtualized services like mobile edge computing more efficiently.

## III. TYPES OF MACHINE LEARNING AND BIG DATA ANALYTICS

There is a significant differnce between Data Analytics and ML. Data analytics has evolved from descriptive analytics to diagnostic analytics to predictive analytics and is excelling towards presciptive analytics [1]. Machine learning is predictive analytics and prescriptive analytics, meaning its models are capable to taking decisions. These decisions are based on parameters set by network operators like system constraints, spectrum, power, etc. The paper [1] describes the data sources and perspectives which are exploited to design and optimize the networks

*1) Subscriber Profile:* This attribute consists of device profile, service level agreement (SLA), QoS, behavioural profile of the user, etc [1].

*2) Subscriber Perspective:* This attribute associates network operator's network activity with the subscriber profile providing a user centric network parameter.

*3) RAN Perspective :* This attribute is associated with data about the user RAN, i.e., signal strength, error codes, available networks etc. This helps in spacial laying of base stations (BS) and signalling patterns to determine optimum RAN quality from users perspective.

*4) Subscriber Mobility Patterns:* This attribute is associated with spatio-temporal trajectory of the user.

*5) Radio Environment Map:* This attribute is linked to the spatio-temporal radio atmosphere, which is described by radio signal attenuation. This attribute along with Subscriber Mobility Patterns facilitate the prediction of average channel gains. [1]

*6) Traffic Profile:* This attribute is linked to the temporal traffic trace, BS spatial deployment and BS characteristics to help predict future network congestion and traffic.

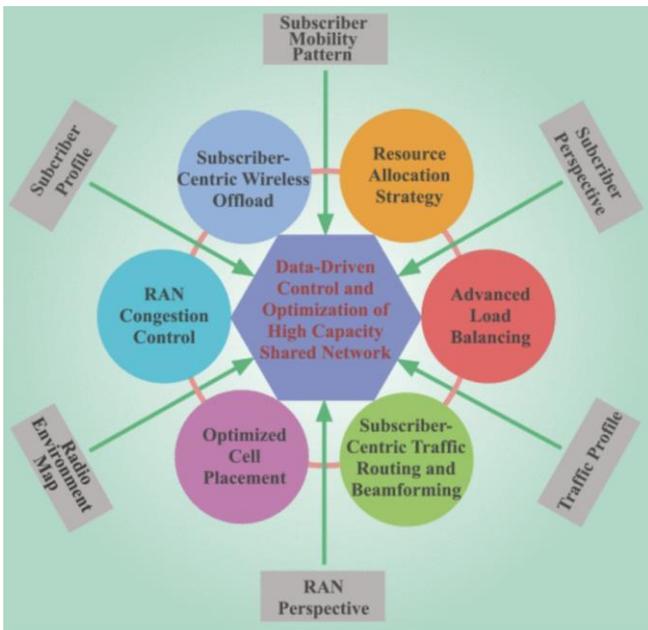

Fig. 1. Data Attributes and their use in Control and Optimisation [1]

Paper [2] describes different categories of ML and discusses the usage of each category in wireless communication technologies. It divides the categories according to level of supervision needed to train the ML model.

## A. Supervised Learning

In supervised learning the programmer feeds the model with sample data that have a known outcome. Each Sample value is mapped to one outcome. The idea is to train the model by feeding in Sample inputs along with known outputs, then use the trained model to predict outputs of new samples. This type of ML Models are beneficial for applications that can access large amounts of data to train their algorithms [2].

This type of ML is used when deploying 5G, LTE small cells to cope with increasing traffic demands. These algorithms gather radio performance measurements such as path-loss and throughput for particular frequencies and bandwidth settings from the cells, [2] and adjust the parameters using ML based alogrithms to predict the performance that a user will experience. This type of machine learning is used to create path loss prediction models, to infer unobservable channel state information, to model and approximate objective functions for link budget and propagation loss for next generation wireless networks [2] and more.

## B. Unsupervised Learning

The data used to train the model is an unlabled collection of different features, and the system attempts to discover subgroups with similar characteristics among the variables without human intervention [2].Algorithms such as K-means clustering, and generative deep neural networks are beneficial when grouping edge computing devices in an network. The authors of the paper [2] suggest that this type of ML is useful for anomaly and fault detection, to store data in clusters in data centers to reduce unnecessary data travel among distributed storage systems, to reduce latency by clustering fog nodes to automatically decide which low power node must be upgraded to a high power node, for data-driven resource management for ultra-dense small cells and more.

## C. Reinforcement Learning

Wireless networks always operate in probabilistic environments. The authors of the paper [2] suggest that the system parameters and dynamics can be modelled using a markov decision process (MDP), to obtain the most efficient and optimum working of the network. A reinforcement learning (RL) model interacts with a network environment, taking decisions at regular intervals of time. The model must choose an action $a$ at every decision point that is available at the current state $s$, to which the system must give a positive or a negative score $R_{(s,a)}$ and then move to a new state. According to the markov property, the state transition probablity $P_{s^t(s,a)}$ is independent of all previous states and actions [2]. Since we need to optimise this decision making process the RL model must keep track of all previous probabilities and their results to help make the most optimum decision at any decision point. The reference [2] denotes the best decision at any point as $\pi^*$. The value return at any arbitrary decision point is denoted via $V^n$. Hence the paper suggests that the best possible outcome value $V^{n^*}$ can predited by the RL model if the MDP is known.

This method can also be used if the MDP is unknown, where the RL model will act as a trial and error learner.

We see that because of the adaptability of RL to change in environment, it can be used for admission control, load balancing, mobility management, resource management, and dynamic channel selection (DSA) [2]. In [2], the authors highlight the usage of Q-learning based techniques for optimising cell range extension (CRE) bias values for each device, along with QoE-based handover decision for HetNets, user scheduling in small cells for energy harvest, proactive resource allocation in LTE-U networks and Jamming resilient control channel in CRNs.

## IV. POTENTIAL OF MACHINE LEARNING IN WIRELESS COMMUNICATION

The refernce [2] studies the potentials of ML in 5G by categorising the different 5G requirements, wherein each category focuses on a different subset of requirements and applications.

### A. Enhanced Mobile Broadband (eMBB)

5G aims to increase data rates from 1Gb/s in 4G to 20Gb/s to support high demand uses. This is achieved by harnessing significant amounts of spectrum resources in mm, cm wave ranges. 5G will also massive multiple input multiple output (MIMO) to provide an increased spectral efficiency [2]. ML will contribute to enhance MIMO by using deep neural networks for channel estimation and direction of arrival (DOA). The paper [2] also suggests that ML contributes in classification of the channel state information (CSI) to select optimum antenna indices. ML is also used in clustering small base stations (SBS) to allow more spectrum resource reuse which helps in more bandwidth per user. It is also used to alleviate the backhaul link congestions by determining the most popular contents at network edge and caching them. To increase dynamic spectral access, RL is used to reduce negative consequences of spectrum sharing on PAL nodes and to access the spectrum opportunistically.

### B. Massive Machine Type Communications (m-MTC)

This deals with the exponentially increasing connectivity demand. This means that there is a sharp increase in the density of devices. The challenge arises for 5G to be able to connect to all these devices, while meeting limited bandwidth and transmission reception points constraints. Q-based learning algorithms have been proposed to enable users to select their serving BS faster by exploiting local data and learning the outcomes of nearest neighbours [2]. ML is been proposed to counter congestions by enabling caching as discussed previously. Linear regression and decision trees have been proposed by the authors of reference [3] to optimise energy harvesting, by defining scheduling policies.

### C. Ultra Reliable Low Latency Communication (URLLC)

5G is meant to reduce latency and provide ultra reliable communication to enable many applications like remote surgery, autonomous driving vehicle to vehicle communication. It is crucial that the requirement of low latency is met with stringent parameters. The authors of paper [2] suggest that URLLC will rely on anticipatory network management, which is capable of predicting future network requirements and reacting accordingly. Latency increases with increase in distance between end device and the servers, hence storing all data in remote servers is not optimal. Peak traffic demands can be substantially reduced by proactively serving predictable user demands via caching at BSs and users' devices. Researchers have obtained significant improvements in the context of edge caching just by applying off-the-shelf ML algorithms, such as k -means clustering and non-parametric Bayesian learning [2].

To achieve maximum QoS in mobile situations, the authors of paper [2] suggest optimal identification of beam forming vectors. ML models can use the uplink pilot signal received at the terminal BSs, and learn the implicit mapping function relating to the environment setup to predict and coordinate beamforming vectors at the BSs [2]

## V. MACHINE LEARNING IN mmWAVE

mmWave plays a big role in bringing 5G to users, by providing large bandwidth and high data rates. As discussed previously, 5G requires deployment of SBSs to help alleviate load on the network. This enables mmWave to be used as a medium of transmission as 5G can capitalise on the bandwidth available at 60GHz and its short range propogation characteristics. This helps facilitate network densification. The reference [3] suggests that user induced channel effects vary significantly with user activity and equipment being used, and are more prominent at mmWave frequencies. The authors of paper [3] suggest that networks must evolve to utilise network radio resource management (RMM) optimisation techniques based on identifying use cases of the user along with user channel induced effects. The paper suggests that the use of wearable sensors is often seen in literature, which helps recognise common UE and use cases.

The authors of the paper suggest a combined supervised and unsupervised that is able to recognise UEs for indoor scenarios. The authors have conducted an extensive mmWave measurement campaign and have laid out a comprehensive methodology to extract small scale fading features, and have compared the classification accuracy and performance of different classifiers, namely multi-SVM, decision trees, Naive Bayes, and Ensemble learning [3].

### A. Methodology

The authors define three common use cases and measure the received signal strength (RSS) for Line of Sight (LOS), Quasi of Line of Sight (QLOS), and Non line of sight (NLOS) parameters, in terms of the orientation angle relative to the direct geometric path to the access point (AP). To extract the small scale fading for analysis, the authors remove path loss and large scale fading in the preprocessing of data, by applying a low pass filter to raw RSS data. To identify whether initial UE use case is either static or in motion, k-means algorithm was proposed by the authors. The number of clusters was set to 2 and variance of small scale fading is fed as input to the algorithm.

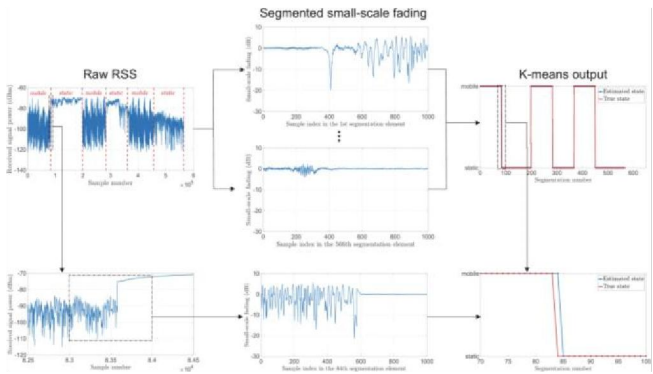

Fig. 2. K-Means preprocessing output example [3]

The authors were able to predict the state of user to an accuracy of 99.8%. The error is dominant in the non-static use case but the authors believe that this error is unavoidable. The low pass filter with correct critical fequency is then applied on the RSS segment, using the results obtained by K-Means. Hence the corresponding small scale fading is obtained as a vector $S_n$.

$$S_n = [s_1, s_2...s_n] \quad (1)$$

$s_n$ is the small scale fading of the n-th segment.

Once the small scale fading was obtained by the authors, they were able to extract 6 statistical features which were then analysed. The 6 features were *Variance, Rice Factor, Nakagami Parameter, Channel Coherence time, AFD, LCR*. The authors noticed that the range of each feature varried widely, hence the values were scaled before being fed into the classifier. A 10-fold cross validation was performed by the authors.

### B. Results

We see that Decision Tree classifier outperforms all others with an accuracy of 100% [3]. The authors suggest that these results are obtained as they authors adopted in the decision tree classifier were highly correlated. It is observed that the results for mobile activity classification are better than those for static activity. The authors suggest the reason for this is that the small scale fading is highly correlated to the mobile activity UE use case. The errors in static case also arise as the movements of the users such as breathing cannot be neglected at mmWave frequencies.

It was determined that a constrained feature set had more impact on user orientation classification with an accuracy of 97.6%-100%. This clearly demonstrates that a UE use case can be identified with precision using RSS Based Classification in indoor mmWave, thus implicating that ML has holds a significant amount of benefits in terms of implementation of mmWave communication techonology.

## VI. CONCLUSION

We must consider that next-generation networks will rely heavily on data. Hence network operators must employ advanced ML for efficient operation, control and optimisation. It is clear that ML will play an important role in implementing 5G, even though network operators are still not sure about implementing ML into their networks. Paper [3] shows the power of ML by classifying the UE use case, which can help in prediction of data congestions, optimisation of QoS and QoE. However one must be careful that ML cannot be applied everywhere due to complexity, cost and latency introduced due to ML. Hence it is important the future researchers must make careful evalutations of trade off between increasing the accuracy of a wireless system by employing an ML based algorithms and the interpretebility of the model and drawbacks involved.

## ACKNOWLEDGMENT

I would like to thank my professor Dr. Naveen Gupta for his constant guidance and support. Without him, this report would not have been possible. I would also like to thank him for introducing me to the concept of literature review and research in the field of Machine learning for Wireless Communication.

## REFERENCES

[1] M. G. Kibria, K. Nguyen, G. P. Villardi, O. Zhao, K. Ishizu and F. Kojima, "Big Data Analytics, Machine Learning, and Artificial Intelligence in Next-Generation Wireless Networks," in IEEE Access, vol. 6, pp. 32328-32338, 2018, doi: 10.1109/ACCESS.2018.2837692.
[2] M. E. Morocho-Cayamcela, H. Lee and W. Lim, "Machine Learning for 5G/B5G Mobile and Wireless Communications: Potential, Limitations, and Future Directions," in IEEE Access, vol. 7, pp. 137184-137206, 2019, doi: 10.1109/ACCESS.2019.2942390.
[3] L. Zhang, Y. Hua, S. L. Cotton, S. K. Yoo, C. R. C. M. Da Silva and W. G. Scanlon, "An RSS-Based Classification of User Equipment Usage in Indoor Millimeter Wave Wireless Networks Using Machine Learning," in IEEE Access, vol. 8, pp. 14928-14943, 2020, doi: 10.1109/ACCESS.2020.2966123.